# Anisotropic Magneto Resistance and Planar Hall Effect at the LaAlO$_3$/SrTiO$_3$ Heterointerfaces: Effect of Carrier Confinement on Magnetic Interactions

A. Annadi,[1,2] Z. Huang,[1] K. Gopinadhan,[1,3] X. Renshaw Wang,[1,2] A. Srivastava,[1,2] Z. Q. Liu,[1,2] H. Ma,[1,2] T. Sarkar,[1,2] T. Venkatesan,[1,2,3] Ariando[1,2,a]

[1]*NUSNNI-Nanocore, National University of Singapore, 117411 Singapore,*

[2]*Department of Physics, National University of Singapore, 117542 Singapore,*

[3]*Department of Electrical and Computer Engineering, National University of Singapore, 117576 Singapore*

[a)]*ariando@nus.edu.sg*

**The confinement of the two dimensional electron gas (2DEG), preferential occupancy of the Ti 3d orbital and strong spin-orbit coupling at the LaAlO$_3$/SrTiO$_3$ interface play a significant role in its emerging properties. Here we report a fourfold oscillation in the anisotropic magneto resistance (AMR) and the observation of planar Hall effect (PHE) at the LaAlO$_3$/SrTiO$_3$ heterointerface. We evaluate the carrier confinement effects on the AMR and find that the fourfold oscillation appears only for the case of 2DEG system while it is twofold for the 3D system. As the fourfold oscillation fits well to the phenomenological model for a cubic symmetry system, we attribute this oscillation to the anisotropy in the magnetic scattering arising from the interaction of electrons with the localized magnetic moments coupled to the crystal symmetry. The AMR behavior is further found to be sensitive to applied gate electric field, emphasizing the significance of spin-orbit coupling at the interface. These confinement effects suggest that the magnetic interactions are predominant at the interface, and the gate electric field modulation of AMR suggest the possible gate tunable magnetic interactions in these systems. The observed large PHE**

**further indicates that the in plane nature of magnetic ordering arises from the in-plane Ti $3d_{xy}$ orbitals.**

The 2DEG observed at the LaAlO$_3$/SrTiO$_3$ interface has shown to exhibit multi functionalities[1-3] in particular recently the co-existence of magnetism and superconductivity.[4,5] The existence of magnetic ordering is demonstrated via several ways such as electronic phase separation,[6] scanning SQUID imaging,[5] torque magnetometry[7] and magneto transport measurements[8,9] at low temperatures. At the LaAlO$_3$/SrTiO$_3$ interface, the electrons occupy Ti $3d$ orbitals of the SrTiO$_3$. The conductivity arises through the itinerant electrons of Ti $3d$ orbitals, while the origin of magnetic ordering is attributed to the localized Ti $3d$ orbitals. Further, owing to the preferential occupancy, particularly to the Ti $3d_{xy}$ orbitals at the interface, various reports have presumed that the magnetization predominantly is in the plane of interface. The LaAlO$_3$/SrTiO$_3$ interface is also shown to exhibit strong spin–orbit coupling[10] due to the broken symmetry near the interface. In this scenario, the confinement of the electron gas at the interface is also expected to influence its orbital occupancy and spin-orbit interaction, and thus its magnetic property. However the interplay between these magnetic phenomena and the carrier confinement is less understood. A natural way to further probe this is by investigating the in plane magneto resistance, *i.e.*, anisotropic magneto resistance (AMR) and planar Hall effect (PHE), which directly relate to the magnetic scattering of itinerant electrons associated with the spin-orbit interaction. In this report, we present a comprehensive study of AMR and PHE on the LaAlO$_3$/SrTiO$_3$ based heterointerfaces with respect to the confinement of the electron gas. We find that a fourfold oscillation in AMR is observed only for confined 2DEGs, while a twofold oscillation is observed for the 3D case. Furthermore, the AMR magnitude is found to depend on the magnitude of the applied magnetic field and temperature. The obtained results are interpreted in terms of the magnetic ordering and spin-orbit interaction coupled to the symmetry of the system.

The LaAlO$_3$/SrTiO$_3$ samples with a LaAlO$_3$ thickness of 8 unit cells (uc) on TiO$_2$ terminated SrTiO$_3$ (001) substrates were grown by pulsed laser deposition (PLD) with *in-situ* reflection high energy electron diffraction (RHEED). For investigation of carrier confinement effects, various samples were grown in different oxygen (O$_2$) partial pressure ranging from $1\times10^{-5}$ to $1\times10^{-3}$ Torr. It has been shown that the O$_2$ pressure during the sample growth can be used to control the dimensionality of the electron gas.[8,11,12] For high O$_2$ pressure ($>1\times10^{-4}$ Torr) grown samples, the carrier density, $n_s$, is found to be typically of the order of $\sim10^{13}$-$10^{14}$ cm$^{-2}$ with a 2D confined conducting channel, whereas for samples grown at low O$_2$ pressures ($\leq1\times10^{-5}$ Torr), the $n_s$ is of the order of $10^{16}$ cm$^{-2}$ and has a 3D like conducting channel. Further, we also fabricated LaAlO$_3$/SrTiO$_3$ heterointerfaces on NdGaO$_3$ (110) substrate, a technique employed recently to grow precise 2DEG.[13,14] In this last configuration the dimensionality of electron gas can be controlled by thickness of SrTiO$_3$. All electrical transport characterizations were carried out by physical property measurement system (PPMS) assisted with rotator which enables to perform angle dependence measurements. AMR and PHE measurements were performed under various parameters including magnetic field and temperature. Further, in-plane MR, gate voltage and current dependence of AMR were also performed to clarify the misalignment and magnetic field wobbling issues in these types of measurements (supplementary materials). Carrier density ($n_s$) for the various samples is extracted from conventional hall measurements.

Figure 1a shows the temperature dependence of the sheet resistance, $R_s(T)$, for the samples grown at various O$_2$ pressures. The high pressure samples show the typical metallic behavior with higher $R_s$ when compared to the ones grown at low O$_2$ pressure. The O$_2$ pressure dependence of $R_s(T)$ shown here is consistent with earlier reports.[8, 12] Figure 1b shows the $n_s$ dependence with temperature for the corresponding samples. For the high $O_2$ pressure samples, the $n_s$ is of the order of $6\times10^{13}$ cm$^{-2}$ at 300 K and $2.5\times10^{13}$ cm$^{-2}$ at 2 K, whereas for the low pressure sample the $n_s$ is about $8\times10^{15}$ cm$^{-2}$ at 300 K and $7\times10^{15}$ cm$^{-2}$ at 2 K. The origin of large $n_s$ in low $O_2$ pressure samples is due to the creation of oxygen vacancies in SrTiO$_3$. The dimensionality of electron gas based on $n_s$ has been previously investigated through various experiments.[11, 15]

We first investigate the AMR effect on the LaAlO$_3$/SrTiO$_3$ samples grown at high $O_2$ pressure, *i.e.* on the confined 2DEG. In AMR measurement, both the magnetic field (***H***) and current (***I***) are in the plane of the sample, and the angle between ***H*** and ***I*** is varied from 0° to 360°. Here, ***I*** is fixed along the <100> direction (depicted in Fig. 2). Figure 2a shows the AMR measured at 2 K with varying ***H*** from 3 to 9 T on the sample grown at $1\times10^{-4}$ Torr, where the AMR is defined as AMR=$(R(\theta)-R(0)/R(0))\times100\%$, $\theta$ is the angle between ***H*** and ***I***, and $R(0)$ is the resistance when ***H*** and ***I*** are parallel to each other. For the AMR measured at 9 T, a clear fourfold oscillation appears in the AMR with resistance minima appearing at an angle 45° to the principal {100} directions with a repetition interval of 90°. Further, when ***H*** is decreased, a clear decrease in the amplitude of the fourfold oscillation and simultaneously a gradual transformation from fourfold to twofold (with two resistance maxima now) oscillation are observed. At ***H***=3 T the oscillations turn completely into twofold with resistance maxima appear at 90° and 270° with an interval of 180°. A noticeable observation here is the change of the AMR sign from negative for the fourfold oscillation to positive for the twofold oscillation. Figure 2b shows the AMR measured with varying temperature from 2 to 20 K at ***H***=9 T. The fourfold response in the AMR gradually decreases when temperature increases and simultaneously a twofold component emerge. The fourfold oscillation behavior in AMR has been previously reported in magnetic systems such as manganites[16-19] and $Fe_3O_4$.[20] Very recently, a similar kind of signature of AMR is also observed at the $LaAlO_3/SrTiO_3$ interface.[21, 22] Strikingly, all the above mentioned systems possess *d*-orbital character. This suggests that the magnetic interactions arising from the *d*-orbitals is crucial for the observed fourfold oscillation.

A phenomenological model is widely used to quantitatively describe the AMR and PHE in various systems.[23,24, 25] In this model, the resistivity tensor is expressed in terms of the direction of current with respect to the applied magnetic field. The fourfold oscillation behavior in the AMR is suggested to be arising from the contribution of higher order terms in the resistivity tensor relating to the crystal symmetry of the system.[24] For the cubic symmetry system, the variation in resistance with angle (θ) is expected to follow the equation containing direction cosines of higher order given by:

$$R_{XX} = C_0 + C_1 \cos^2(\theta + \theta_c) + C_2 \cos^4(\theta + \theta_c) \quad (1)$$

$$R_{XY} = C_3 \sin 2\theta \quad (2)$$

Here $R_{XX}$ corresponds to the AMR where as $R_{XY}$ is to the PHE. The coefficients $C_1$, $C_2$ and $C_3$ are considered to arise from the uniaxial and cubic components of magnetization, and $C_0$ and $\theta_c$ are additive constants introduced to take the observed asymmetry in the magnitude in the four oscillations into account. Figure 2c shows the fitting for the experimental data by Eq.1, showing a good agreement with the above formula. The further striking observation is the asymmetry in magnitude among four oscillations. If one considers just the crystalline anisotropy then symmetry is expected in these fourfold oscillations. The fact that we see an asymmetry strongly suggests that there is further uniaxial anisotropy present in the system. Such anisotropy with respect to the crystal symmetry probably arises because of Fermi surface reconstruction of the various 3$d$ orbitals at the interface. The good agreement of experimental data with the model implies the predominant role of crystal symmetry on the magnetic interactions at the LaAlO$_3$/SrTiO$_3$ interface.

Further, the twofold oscillation observed at 2 K and 3 T is found to follow a $\sin^2\theta$ dependency. The corresponding $\sin^2\theta$ fit for the twofold oscillation is shown in Fig. 2c which matches the experimental data accurately. The origin of the twofold oscillation can be from the Lorentz scattering of charge carriers which follows the $\sin^2\theta$ dependency. The transformation of the fourfold to twofold oscillations infers that there are two competing components (spin and charge scatterings) for the AMR, and their contribution to AMR depends on parameters such as magnetic field and temperature. The evolution of fourfold oscillation from twofold oscillation for the magnetic field $H > 3$ T implies that a critical filed strength is needed for of the coherent magnetic scattering to overcome the charge scattering. Furthermore, the signature of the AMR starts to diminish for higher temperatures (above 20 K) which could be due to the diminishing of coherence scattering as thermal fluctuations emerge strongly.

To further study the anisotropy we performed the planar Hall effect (PHE) measurement. The PHE is similar to the AMR which arises from magnetic related anisotropy. In PHE measurement, $I$ and $H$ are in the plane of the sample and the angle between $H$ and $I$ is varied between 0° to 360°. Unlike AMR, here the resistivity is measured using Hall geometry, *i.e.*, transverse resistance, $R_{XY}$, (depicted in Fig. 3). Figure 3a shows the $R_{xy}$ measured at $H$=9 T with respect to angle (θ) by varying the temperature. The maxima (which are positive) and minima (which are negative) in $R_{XY}$ appear at 45° to the principal {010} directions with 90° interval. In general,

PHE contribution to $R_{XY}$ is expected to follow sin 2θ behavior.[26, 27] Figure 3b shows a fit to sin 2θ for the PHE measured at 2 K and 9 T, which is in good agreement with the formula. The sign change of Hall signal with angle is a characteristic of PHE. Considering the cubic symmetry of the system[24, 25], the AMR can be fourfold and/or two fold, but the PHE is only twofold, which are in fact what we observed in our AHE and PHE (Figs. 2 and 3, respectively). The large planar Hall signal observed here infers the strong in plane component for the magnetic ordering at this interface.

From the results presented above it is shown that the samples having 2D confined electron gas show strong anisotropy with evolution of fourfold oscillations in AMR (a similar oscillation behavior in AMR is also observed for the LaAlO$_3$/SrTiO$_3$ sample grown at the 1×10$^{-3}$ Torr, shown in supplementary material). It is noted here that in this O$_2$ pressure growth regime (where we observed the fourfold oscillation in AMR) various reports have demonstrated the magnetism at this interface using different techniques, although the strength of the magnetic interaction can be further dependent on several sample growth parameters.

To investigate the effect of the confinement of the electron gas, we performed AMR measurements on the LaAlO$_3$/SrTiO$_3$ sample grown at low pressure (1×10$^{-5}$ Torr) which has 3D like conducting channel. Figure 4a shows the AMR measured at 9 T with varying temperatures for the corresponding sample. Remarkably, in contrast to high pressure samples, it shows only twofold oscillation throughout the temperature range of 2-20 K. However, an anti-symmetric behavior in AMR is noticed between 0° to 180° and 180° to 360° for *I* direction with respect to *H* at low temperatures, and it gradually decreases with increasing temperature suggesting the intrinsic origin for this anti symmetry. We attribute this anti-symmetry to the difference in inter boundary scattering between the two sides of the conducting channel in the 3D case, *i.e.*, the interface scattering is stronger at the interface between the LaAlO$_3$ and SrTiO$_3$ while it is negligible deeper in the SrTiO$_3$ side. While the electrons scattered toward the interface encounter the sharp boundary to the LaAlO$_3$ over layer, those scattered into the SrTiO$_3$ side will see a relatively graded boundary due to the 3D nature of the conducting channel. The decrease in anti-symmetry with increase in temperature can be understood as at elevated temperatures the radius of the orbital path becomes smaller than the 3D channel width, thus the electron path is governed entirely by Lorentz scattering as it can be seen from the data that at 20 K the AMR follows $Sin^2\theta$ behavior. A similar twofold oscillation behavior was also reported in the AMR of the Ar-irradiated SrTiO$_3$[28] which has the 3D conducting channel (with $n_s$ ~1×10$^{17}$ cm$^{-2}$). To validate the confinement effects, a conducting heterointerface with LaAlO$_3$ (15 uc)/SrTiO$_3$ (8 uc) is grown on NdGaO$_3$ (110) substrate.[19] In this case the electron gas is precisely confined to 8 uc (*i.e.* thickness of SrTiO$_3$). Figure 4b shows the AMR measured at 9 T with varying temperature (2 - 20 K) for the corresponding sample. Evidently, here also a fourfold oscillation is observed and further the magnitude of AMR gradually diminishes as the temperature increases. As the conducting channel is confined to within the 8 uc, thus the strong interface effects are also expected here similar to the high pressure grown LaAlO$_3$/SrTiO$_3$ samples.

At the LaAlO$_3$/SrTiO$_3$ interfaces, the electric field applied to interface through back gate voltage was shown to tune the spin-orbit interaction[10] and also the carrier density modulation[29] which is further interpreted to influence the preferential *d*-orbital filling[30]. To examine the influence of spin-orbit interaction on AMR, we further performed a gate electric field dependence of the AMR on the sample grown at 1×10$^{-4}$ Torr. Figure 5 shows the gate electric field dependence of the AMR behavior measured at 3 K and 9 T. The strength of the fourfold oscillation for a positive +30 V gate voltage is enhanced when compared to the +0 V case. On the other hand, the fourfold oscillation turns in to a twofold oscillation for the -30 V gate voltage, showing a strong gate electric field (polarity) dependence of AMR at these interfaces. The modulation of AMR with electric field effects further confirms that the spin orbit coupling at the interface has a key role governing the magnetic interactions and this provides an opportunity to tune the magnetic interactions by electric field.

From the above observations, it is evident that the confinement of the electron gas and the spin-orbit interaction strength at the interface are very crucial to observe the fourfold oscillation in AMR. For the case of confined 2DEG, numerous experimental and theoretical reports illustrated that, near the LaAlO$_3$/SrTiO$_3$ interface, the Ti 3$d$-orbitals undergo crystal field induced splitting which leads to a preferential filling[31-33] of 3$d$-orbitals. It is suggested that the occupation is predominantly be the 3$d_{xy}$ orbitals in the first few layers of SrTiO$_3$[31,32] from the interface which is supported by the spectroscopic experiments.[33] Further, the fraction of occupied 3$d_{xy}$ orbitals (Ti$^{3+}$) near the interface is expected to have localized character due to lattice coupling and/or disorder which gives rise to magnetic ordering. Therefore the interaction of itinerant electrons to the interface localized in plane 3$d_{xy}$ orbitals in presence of external magnetic field would be the prime source of the observed coherent magnetic scattering. Recently, M. Trushin *et al.* [34] discussed the combined effects of spin orbit coupling and symmetry of the system on AMR is by considering the interplay between spin orbit coupling and polarized magnetic moments,[34] which also further analyzed the evolution of crystalline AMR with respect to the orientation of current and magnetization in spin -orbit coupling scenario.

The experimental observations such as the fourfold oscillations, crystalline AMR (with uniaxial anisotropy among the oscillations) and gate electric field dependence of AMR via tuning the spin-orbit coupling as shown here could support the above scenario at these interfaces. Thus we suggest that the interplay between spin-orbit coupling associated with the symmetry and magnetism would be the key to observe this characteristic AMR behavior. Referring to the 3D case, the $n_s$ is of the order of $10^{16}$ cm$^{-2}$, and the special extension of carriers is deeper from the interface into the SrTiO$_3$ side. The consequence of 3D like conducting channel are; the interface crystal field would be potentially screened by this large $n_s$ due to the strong Coulomb interactions, as a result the *3d*-orbitals would become degenerate in SrTiO$_3$, and the strength of the spin-orbit coupling is weaken along with the depth of the electron gas from the interface, and moreover the large $n_s$ in this case may diminish also the magnetic ordering near the interface as it can reduce the carrier localization by minimized interface effects. Thus we suggest that due to the weak magnetic interactions among carriers no fourfold oscillation observed in the AMR in 3D case. The results presented above establish that the fourfold oscillation appear only for the confined 2DEG, and indicates that the magnetic ordering is predominant near the interface and would be weaken far from the interface.

In summary, we have shown the AMR and PHE in the LaAlO$_3$/SrTiO$_3$ system. A fourfold oscillation behavior in the AMR is observed for the confined 2DEG, and a twofold oscillation observed for the 3D case. The observation of this behavior only for the confined 2DEG case infers that the magnetic interactions are predominant at the interface. The modulation of AMR with gate electric field effects further provides an opportunity of tuning magnetic interactions with gate electric fields via tuning the spin-orbit coupling. The observed PHE suggests that the predominant component of the magnetization is in the plane of the samples. The origin of the fourfold oscillation primarily would be arising from strength of the magnetic interaction of itinerant electrons with localized magnetic moments associated with spin-orbit interaction and coupled to the crystal symmetry. The AMR and PHE measurements would be very useful as a probe for the magnetic interactions in low dimensional systems.

**Acknowledgement**

We thank the National Research Foundation (NRF) Singapore under the Competitive Research Program (CRP) "Tailoring Oxide Electronics by Atomic Control" NRF2008NRF-CRP002-024, National University of Singapore (NUS) cross-faculty grant and FRC (ARF Grant No. R-144-000-278-112) for financial support.

**Figure Captions:**

**Figure 1:** (a) Temperature dependence of Sheet resistance, $R_s$, of the LaAlO$_3$/SrTiO$_3$ samples grown at various pressures. (b) Carrier density, $n_s$, variation with temperature for corresponding samples.

**Figure 2:** AMR measured for the LaAlO$_3$/SrTiO$_3$ interface prepared at 1x10$^{-4}$ Torr, (a) various temperatures, and (b) magnetic fields. (c) A phenomenological model formula fit to the AMR obtained at 2 K and 9 T (d) $\sin^2\theta$ fit for the AMR at 2 K and 3T.

**Figure 3:** (a) PHE measured for the LaAlO$_3$/SrTiO$_3$ interface prepared at 1x10$^{-4}$ Torr at various temperatures. (b) $\sin 2\theta$ fit for the PHE obtained at 2 K and 9T.

**Figure 4:** (a) AMR measured for the LaAlO$_3$/SrTiO$_3$ interface prepared at 1x10$^{-5}$ Torr ( 3D conducting channel ) at various temperatures at 9 T. (b) AMR measured for the LaAlO$_3$(15 uc)/SrTiO$_3$(8 uc) / NdGaO$_3$ (110) hetero structure ( 2D conducting channel) at various temperatures at 9 T.

**Figure 5:** (a) AMR measured for the LaAlO$_3$/SrTiO$_3$ interface prepared at 1x10$^{-4}$ Torr with various back gate voltages at 3 K and 9 T.

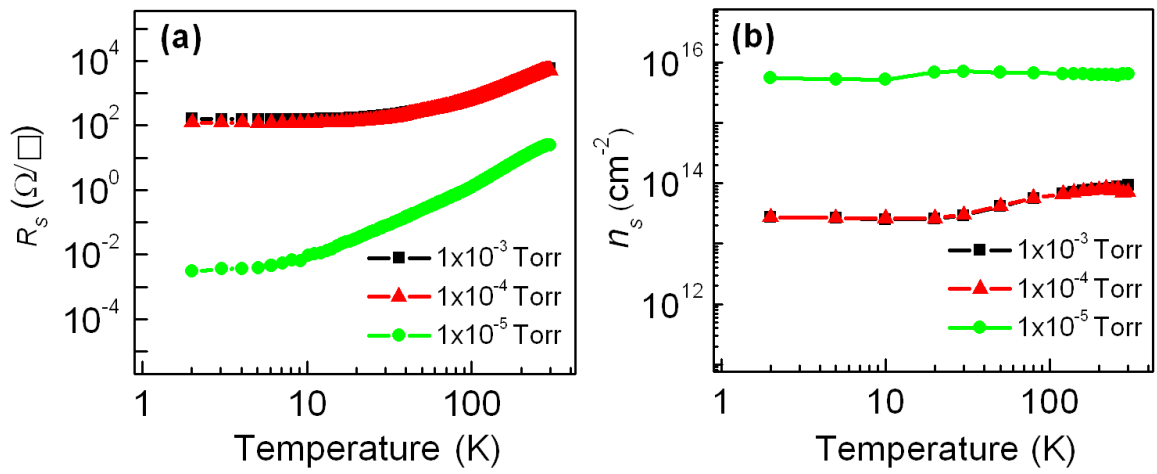

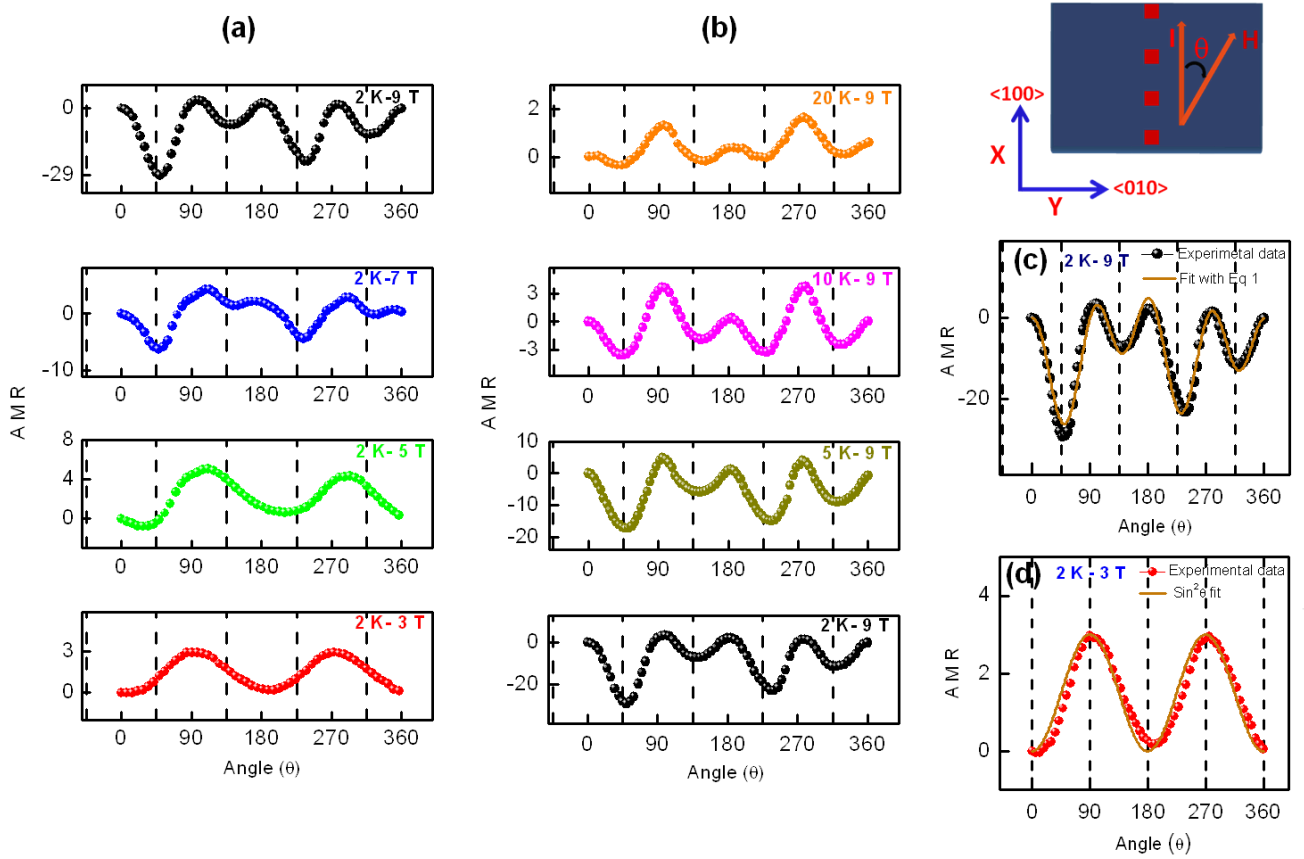

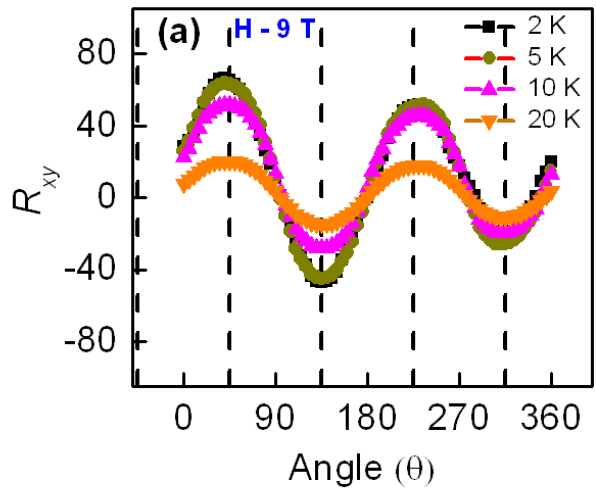

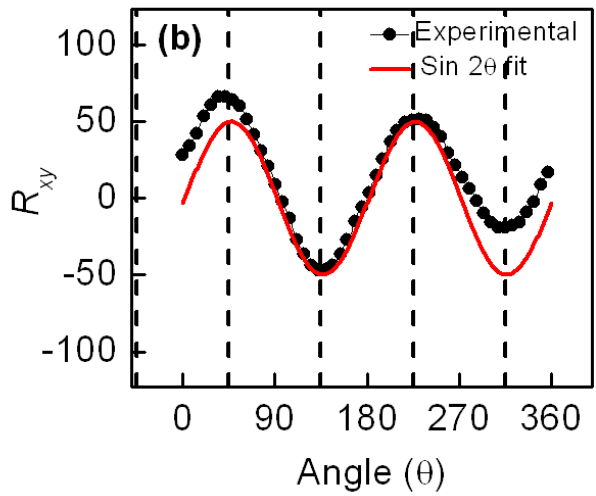

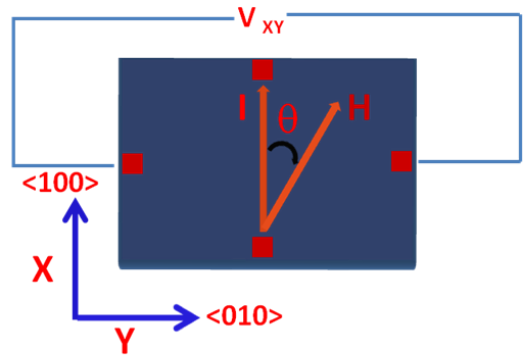

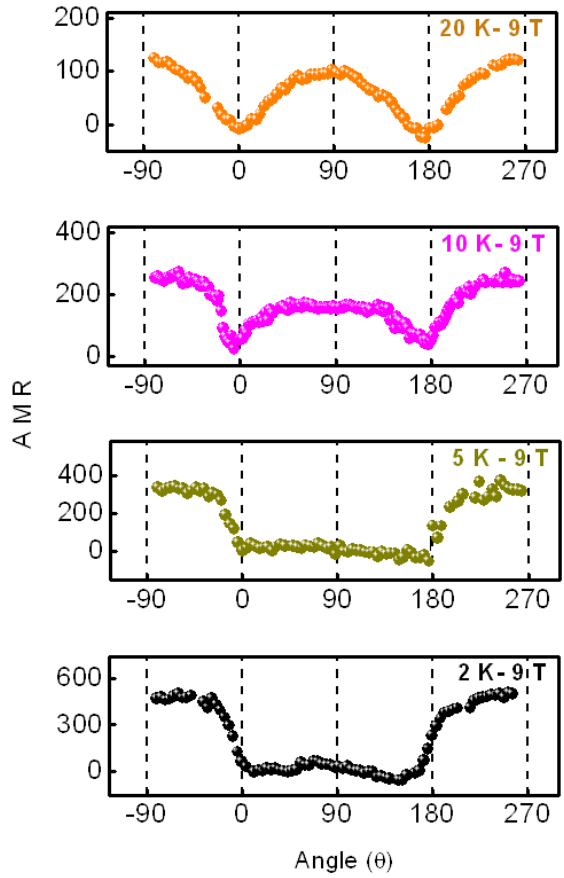
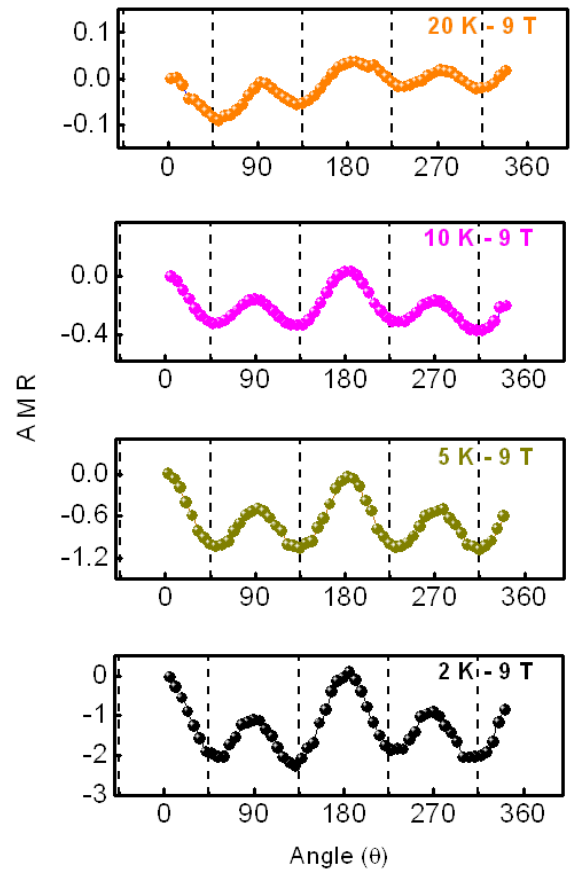

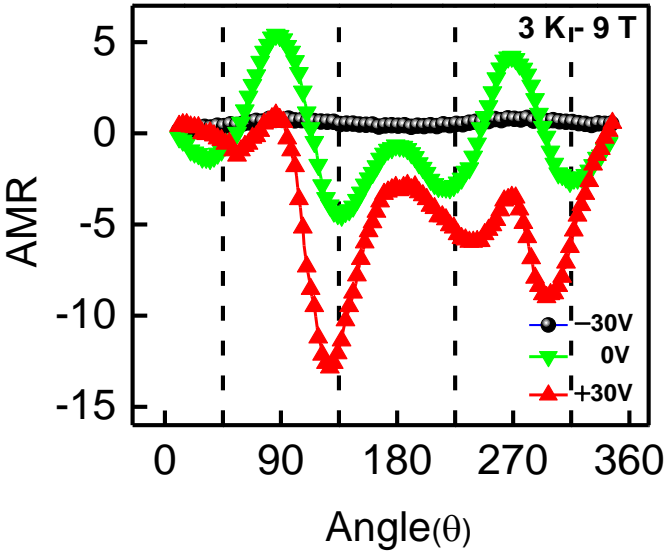

# Supplementary Materials for

# Anisotropic Magneto Resistance and Planar Hall Effect at the LaAlO$_3$/SrTiO$_3$ Heterointerfaces: Effect of Carrier Confinement on Magnetic Interactions

A. Annadi,[1,2] Z. Huang,[1] K. Gopinadhan,[1,3] X. Wang,[1,2] A. Srivastava,[1,2] Z. Q. Liu,[1,2] H. Ma,[1,2] T. Sarkar,[1,2] T. Venkatesan,[1,2,3] Ariando[1,2,a]

*[1]NUSNNI-Nanocore, National University of Singapore, 117411 Singapore,*

*[2]Department of Physics, National University of Singapore, 117542 Singapore,*

*[3]Department of Electrical and Computer Engineering, National University of Singapore, 117576 Singapore*

*[a]ariando@nus.edu.sg*

## 1. Magneto transport for LaAlO$_3$/SrTiO$_3$ interface sample grown at 1x10$^{-4}$ Torr.

Figure S1a shows the temperature dependence of R$_s$ with in-plane magnetic field (0 and 9 T). At low temperatures at about 30 K, the R$_s$ for 9 T case start to deviate from R$_s$ for without magnetic field case, implies a negative magneto resistance at low temperatures for the in-plane magnetic field case. The negative magneto resistance in in-plane mode is an indicative of magnetic scattering and this observation is consistence with earlier reports[1]. Further, the deviation temperature of 30 K is also co-insides with the temperature scale of clear observation of AMR as shown in the main text (Figure 2). Figure S1b shows in-plane magneto resistance measured at 2 K and up to 9 T for fixed angles θ = 0º, 90º (θ is the angle between current and in plane magnetic field). The data shows a symmetric MR behavior with the magnetic field direction suggest absence of any component of normal Hall contribution to the measurements due to wobbling of magnetic field perpendicular components. In addition to that there appear a difference in critical field like behavior above which the magneto resistance start to change rapidly between the data of θ = 0º and 90º. The critical field strength is about - 3 T for θ = 0º and 1.5 T for θ = 90º. The observed higher critical field for θ =0º compared to θ =90º is interesting and could suggests that the magnetic polarization effects are large for the θ =90º.

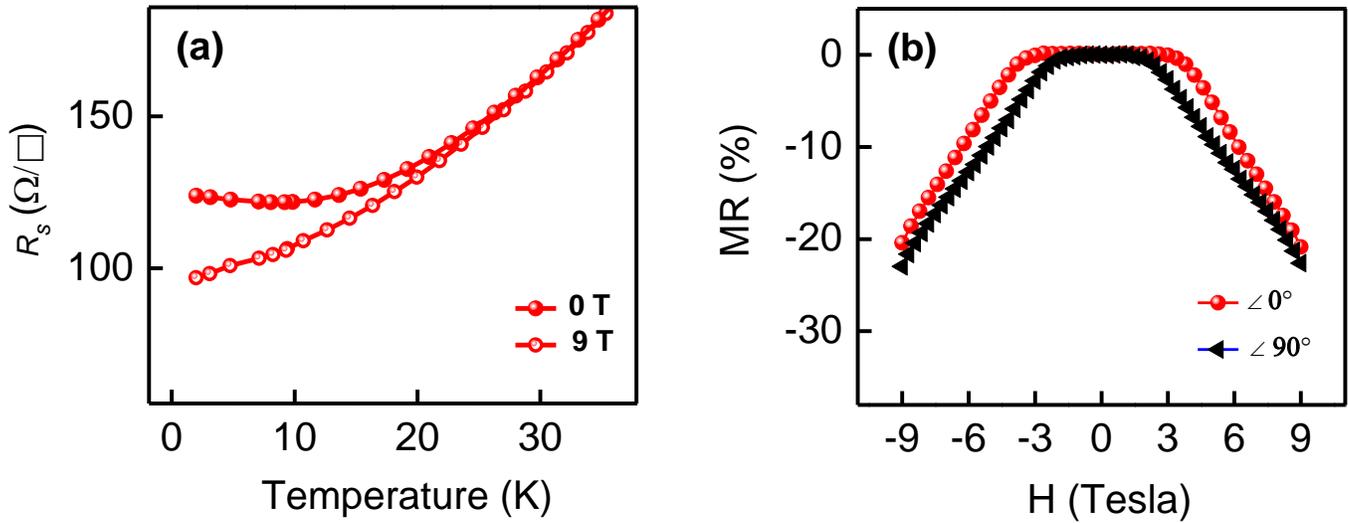

**Fig. S1.** Magneto transport properties of $LaAlO_3/SrTiO_3$ interface prepared at $1 \times 10^{-4}$ Torr, (a) Temperature dependence of $R_s$ with in-plane magnetic field (0, 9 T) and (b) in-plane magneto resistance measured at 2 K and up to 9 T for fixed angles θ = 0º, 90º (θ is the angle between current and in plane magnetic field).

We further perform the AMR measurement with changing the magnitude of applied current. Figure S2 shows the corresponding AMR behavior measured at 2 K and 9 T for the current range of 0.5µA-500 µA. The data clearly shows no dependence of the magnitude of current on fourfold oscillation behavior which explicitly rules out the possibility of contribution from any activated carrier behavior due to the continuous application of magnetic fields and possible thermal effects in the sample to the observed AMR behavior.

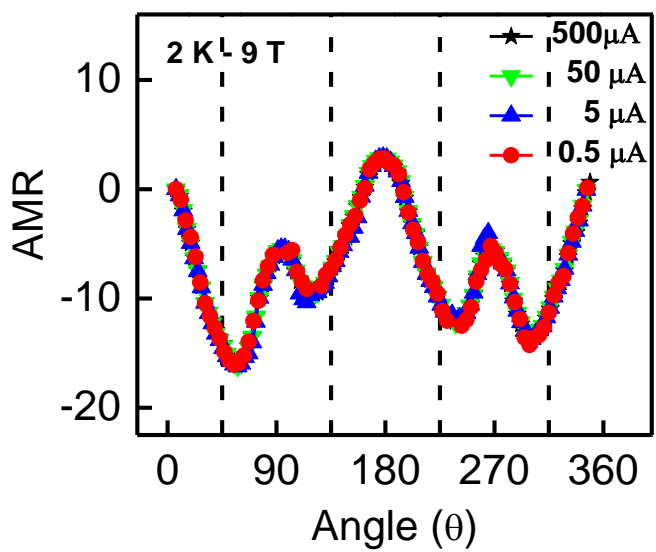

**Fig. S2.** AMR measured at 2 K and 9 T with different magnitudes of current for the LaAlO$_3$/SrTiO$_3$ interface prepared at 1x10$^{-4}$ Torr.

We conclude from the presented data of magneto transport and AMR under various parameters that the data certainly indicates the magnetic origin in the system, and the gate voltage dependence further confirms that the AMR behavior is intrinsic of the system and rules out the misalignment effects.

## 2. AMR measured for the LaAlO$_3$/SrTiO$_3$ interface sample grown at 1x10$^{-3}$ Torr

Figure S3 shows the AMR measured at 9 T with varying temperatures from 2 K to 20 K on the sample grown at 1×10$^{-3}$ Torr. For the AMR measured at 2 K, the resistance minima appearing at an angle 45° to the principal {100} directions however with a repetition interval of 180°. Further, as the temperature increased a clear fourfold oscillation emerges for 10 K and 20 K temperatures with a repetition interval of 90°. The subtle difference between two samples is that at 2 K and 9 T high pressure sample shown a further preferential anisotropy. Note here that though the carrier density of the samples is identical (~2.5×10$^{13}$ cm$^{-2}$ at 2 K) yet the AMR shows distinct differences which can arise from and due to the differences in growth parameters, spin-orbit interaction and electronic ground state in terms of electronic reconstruction at the interface.

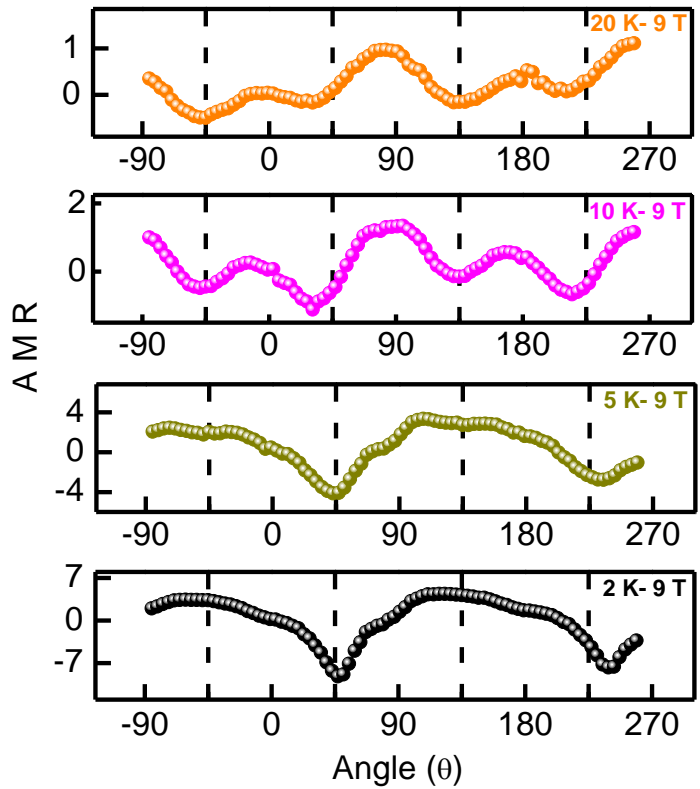

**Fig. S3.** AMR measured at various temperatures for the LaAlO$_3$/SrTiO$_3$ interface sample grown at 1x10$^{-3}$ Torr .

**Experimental details:**

All electrical transport characterizations were carried out by physical property measurement system (PPMS) with the capability to reach temperature of 2 K and magnetic field 9 T respectively. System consist a sample probe where the sample stage can be rotated with a precise control of rotation motion sensor which enable to perform in plane magnetic field measurements. The gate voltage experiments were performed by using $SrTiO_3$ as a gate material[2]. During the gate measurements issues such as carrier trapping and leakage current were taken care.